\documentclass[twocolumn]{aastex631}

\newcommand{\uvot}{UV/optical}

\usepackage{amsmath}

\shorttitle{AGN frequency-resolved lags}    
\shortauthors{Panagiotou et al.}
\begin{document}

\title{Frequency-resolved time lags due to X-ray disk reprocessing in AGN}

\correspondingauthor{Christos Panagiotou}
\email{cpanag@mit.edu}

\author{Christos Panagiotou}
\affiliation{MIT Kavli Institute for Astrophysics and Space Research, Massachusetts Institute of Technology, Cambridge, MA 02139, USA}

\author{Iossif Papadakis}
\affiliation{Department of Physics and Institute of Theoretical and Computational Physics, University of Crete, 71003 Heraklion, Greece}
\affiliation{Institute of Astrophysics, FORTH, Voutes, GR-7110 Heraklion, Greece}

\author{Erin Kara}
\affiliation{MIT Kavli Institute for Astrophysics and Space Research, Massachusetts Institute of Technology, Cambridge, MA 02139, USA}

\author{Marios Papoutsis}
\affiliation{Department of Physics and Institute of Theoretical and Computational Physics, University of Crete, 71003 Heraklion, Greece}
\affiliation{Institute of Astrophysics, FORTH, Voutes, GR-7110 Heraklion, Greece}

\author{Edward M. Cackett }
\affiliation{Department of Physics and Astronomy, Wayne State University, 666 W Hancock Street, Detroit, MI 48201, USA}

\author{Michal Dov\v ciak}
\affiliation{Astronomical Institute of the Academy of Sciences, Bo\v cn\'i II 1401, CZ-14100 Prague, Czech Republic}

\author{Javier A. Garc\'ia}
\affiliation{X-ray Astrophysics Laboratory, NASA Goddard Space Flight Center, Greenbelt, MD 20771, USA}
\affiliation{Cahill Center for Astronomy and Astrophysics, California Institute of Technology, 1200 California Boulevard, Pasadena, CA 91125, USA}

\author{Elias Kammoun}
\affiliation{Cahill Center for Astronomy and Astrophysics, California Institute of Technology, 1200 California Boulevard, Pasadena, CA 91125, USA}

\author{Collin Lewin}
\affiliation{MIT Kavli Institute for Astrophysics and Space Research, Massachusetts Institute of Technology, Cambridge, MA 02139, USA}

\begin{abstract}

% 250 words max/tentative

Over the last years, a number of broadband reverberation mapping campaigns have been conducted to explore the short-term UV and optical variability of nearby AGN. Despite the extensive data collected, the origin of the observed variability is still debated in the literature. Frequency-resolved time lags offer a promising approach to distinguish between different scenarios, as they probe variability on different time scales. In this study, we present the expected frequency-resolved lags resulting from X-ray reprocessing in the accretion disk. The predicted lags are found to feature a general shape that resembles that of observational measurements, while exhibiting strong dependence on various physical parameters. Additionally, we compare our model predictions to observational data for the case of NGC 5548, concluding that the X-ray illumination of the disk can effectively account for the observed frequency-resolved lags and power spectra in a self-consistent way. To date, X-ray disk reprocessing is the only physical model that has successfully reproduced the observed multi-wavelength variability, in both amplitude and time delays, across a range of temporal frequencies.

\end{abstract}

% \keywords{***}

\section{Introduction} 
\label{sec:intro}

Active galactic nuclei (AGN) are among the most luminous celestial sources, powered by the accretion of matter onto a supermassive black hole. Ubiquitous across all redshifts, they play an important role in the evolution of the universe \citep{KormendyHo2013}. Nonetheless, our understanding of these systems, and especially of their inner accretion flow, remains largely incomplete; with their small angular size posing a major limitation in studying these regions.

In an effort to circumvent this limitation, several AGN have been the subject of broadband reverberation mapping campaigns over the past decade \citep[e.g.][]{McHardy2014, Edelson2015, Cackett2020, Kara2021}. These studies aim to leverage the expected correlated variability of AGN emission across different wavebands in order to effectively map the accretion disk surrounding the central black hole. Under the assumption that the X-ray illumination of the disk drives the observed variability, it is expected that emissions at various wavebands will be well correlated, with bluer wavebands exhibiting greater variability and with redder wavebands' variability being delayed \citep[e.g.][]{Cackett2007}.

The aforementioned works have accumulated a wealth of high quality data that can be used to test different scenarios. Initial investigations deduced that the disk variability is indeed well correlated across the UV and optical wavebands \citep[e.g.][]{Edelson2019}, as quantified by the cross correlation function (CCF). On the other hand, the correlation between X-ray and disk variability was found to be typically smaller than initially expected. This result, though, can be naturally explained by the dynamic variability of the X-ray source \citep{Panagiotou2022ccf}. 

The UV/optical CCF time lags were shown to increase with wavelength, following broadly the relation $\tau \propto \lambda ^{4/3}$, which provided direct observational evidence that the disk temperature follows the theoretically expected radial profile, $T(R) \propto R^{-3/4}$ \citep{SS1973}. However, it was originally thought that the measured lags are larger than %naively 
those expected by simple approximated scaling relations \citep[e.g.][]{Fausnaugh2016}, a trend consistent across all the studied sources \citep[see][for a review]{Cackett2021}. Taken at face value, this result called into question the validity of standard accretion theory.

Nevertheless, soon after, \cite{Kammoun2021} showed that the observed lags are fully consistent with theoretical predictions. These authors simulated the expected disk emission assuming its X-ray illumination taking into account all the physical effects at play. They found a general agreement between the observed time lags and their predictions in a sample of sources \citep{Kammoun2021data, Kammoun2023}. Further studies have demonstrated that X-ray reprocessing by the disk can explain both the observed amplitudes of UV/optical variability, as quantified by power spectral densities (PSDs), and the CCF time lags \citep{Panagiotou2020, Panagiotou2022ApJ}. 

Motivated by the initial apparent mismatch between the observed and expected time lags, an alternative scenario was proposed. It was suggested that part of the observed variability may arise from reprocessing in the more extended broad line region (BLR). This was also driven by the presence of a considerably large lag in the $u$ waveband, where contribution from the Balmer continuum is expected to be most significant. Notably, \cite{Cackett2018} analyzed HST light curves of the nearby Seyfert galaxy NGC 4593 and concluded that the observed lags show a discontinuity around 3700 \AA, similar to the shape of the Balmer jump \citep{KoristaGoad2019}.
It should be mentioned though that this $u$ band excess is not equally prominent across all sources \citep[e.g.][]{Edelson2019, Kammoun2021data, Kara2023}.

Physically, one expects contributions to the observed variability due to reprocessing by both the disk and the BLR. The magnitude of each contribution depends on the physical properties of each medium and on the exact geometry of the system, which determines the amount of central variations intercepted by each surface. While these properties are not known a priori, if BLR variability is considerable, it is expected to become important on longer time scales due to its larger spatial extent. Therefore, investigating the observed variability on different time scales presents a promising way for distinguishing between the contributions of these two components.

The frequency-resolved time lags measure the time delay between variations of two wavebands at different temporal frequencies \citep[e.g.][]{Uttley14}, and are thus an ideal tool for such an analysis. The estimation of the frequency-resolved time lags is performed in the Fourier space, necessitating long and well-sampled light curves. This requirement has limited the broader application of this method in optical studies, while it is more commonly used in X-ray timing investigations of AGN \citep[see][and references therein]{Uttley14}.

\cite{Cackett2022} were the first to apply this method in order to measure the UV/optical time lags of NGC 5548 as a function of Fourier frequency across different wavebands, using the light curves from the AGN STORM campaign \citep{DeRosa2015}. In general, the observed frequency-resolved lags were found to increase towards lower frequencies, with redder wavebands featuring larger lags. The shape of the observed time lags was found to be suggestive of significant contribution from the BLR's diffuse continuum emission when compared to a simple disk model. Recent studies by \cite{Lewin2023, Lewin2024} have yielded similar findings for Mrk 335 and Mrk 817, respectively.

In this work, we use a physical model to simulate the response of the accretion disk and we present a suite of predicted frequency-resolved time lags for a variety of physical properties. We begin by detailing the calculation of the model time lags (Sect. \ref{sec:lag_calc}) and discuss their dependence on different physical properties in Sect. \ref{sec:depend}. We then compare our model predictions against the observed time lags of NGC 5548 in Sect. \ref{sec:5548}, showing that disk X-ray reprocessing can explain well the observed frequency-resolved time lags with no need for an additional reprocessing region. We conclude with our main findings in Sect. \ref{sec:conclude}.

\section{Frequency-resolved time lags estimation}
\label{sec:lag_calc}

\subsection{Theoretical definitions}

Let us consider the disk emission in two distinct wavebands A and B. In the case of X-ray disk reprocessing, their flux at any time $t$ may be written as:

\begin{equation}
\label{eq:fa}
    F_A (t) = F_{NT,A} + \int_{-\infty}^\infty F_{\text{X}} (t-t') \cdot \psi_A(t')  dt',
\end{equation}

\noindent and

\begin{equation}
\label{eq:fb}
    F_B (t) = F_{NT,B} + \int_{-\infty}^\infty F_{\text{X}} (t-t') \cdot \psi_B(t')  dt',
\end{equation}

\noindent respectively. $F_{NT}$ denotes the standard accretion disk emission expected in the case of no X-ray illumination \citep{NT73}, assumed to remain constant on the time scales of interest to this work. $F_\text{X}$ stands for the X-ray flux and $\psi$ is the disk response function \citep[e.g.][]{Cackett2007, Kammoun2021}. Assuming that $\{F_A\}$ and $\{F_B\}$ are jointly wide-sense stationary, their cross covariance is defined as:

\begin{equation}
\label{eq:cov}
    R_{AB} (\tau) = E\{[F_A (t) - \mu_A] \cdot [F_B(t + \tau) - \mu_B]\},
\end{equation}

\noindent where $E$ and $\mu$ denote the expectation operator and the mean value of the flux at each waveband, respectively.

Substituting eqs. \ref{eq:fa} and \ref{eq:fb} in eq. \ref{eq:cov}, it is straightforward to show that

\begin{equation}
    R_{AB} (\tau) = \int_{-\infty}^\infty \int_{-\infty}^\infty \psi_A(t') \psi_B(t'') ACF_X (\tau+t'-t'') dt' dt'',
\end{equation}

\noindent where 

\begin{equation}
\label{eq:acf}
    ACF_{X} (\tau) = E\{[F_X (t) - \mu_X] \cdot [F_X(t + \tau) - \mu_X]\}
\end{equation}

\noindent is the autocovariance of the X-ray light curve with an average $\mu_X$. 
In this case, the cross spectrum between A and B, defined as the Fourier transform of their cross covariance, will be given by:

\begin{equation}
\label{eq:crosspectrum}
\begin{split}
h_{AB} (\nu) & \equiv \int_{-\infty}^\infty R_{AB} (\tau) e^{-i 2\pi\nu\tau} d\nu \\
            & = \Gamma_A^* (\nu) \cdot \Gamma_B(\nu) \cdot PSD_X(\nu)
\end{split}
\end{equation}

\noindent where we have used the correlation theorem and the convolution theorem to obtain the last equality. $PSD_X$ denotes the X-ray power spectrum, which by definition is equal to the Fourier transform of the $ACF_{X}$. The function $\Gamma$ corresponds to the so-called transfer function of each waveband at temporal frequency $\nu$, while the asterisk is used to denote the complex conjugate. The transfer function is defined as the Fourier transform of the response function, $\psi$, and its shape determines the connection between the variability of the responding light curve and that of the driving one \citep[e.g.][]{Panagiotou2022ApJ}.

\subsection{Estimation of phase lags in practice}

As it becomes evident from the above, one needs to estimate the response function of the accretion disk in order to calculate the expected frequency-resolved time lags. We used the latest version of the KYNXiltr\footnote{https://projects.asu.cas.cz/dovciak/kynxiltr} model to simulate the disk response following the approach outlined by \cite{Kammoun2023}. In brief, assuming a lamppost geometry for the X-ray source and the X-ray illumination of the accretion disk, this model computes the disk response function at any wavelength for a given set of parameters, such as the accretion rate, the black hole spin, the source inclination etc. Importantly, taking into account all the general relativistic effects, this model estimates explicitly the incident X-ray flux onto any part of the disk and uses it to estimate the local temperature of the disk \citep[see][for a more detailed discussion of the model]{Dovciak2022, Kammoun2023}.

Having estimated the disk response function, $\psi_\lambda[t]$, for a waveband centered at $\lambda$, with width $\Delta \lambda$, and for a set of times \{$t_k$\}, one can then estimate the corresponding transfer function at a given frequency $\nu_j$ as:

\begin{equation}
\label{eq:transfer_compute}
    \Gamma_\lambda[\nu_j] = \frac{\sum_{k=0}^{N-1} \psi_\lambda[t_k] e^{-2\pi i \nu_j t_k} \Delta t}{\Delta \lambda \cdot e^{i\pi \nu_j T_\text{flash}} \text{sinc}(\pi \nu_j T_\text{flash})},
\end{equation}

\noindent where $N$ and $\Delta t$ denote the total number of points and the bin size of the estimated response, respectively. The exponential and sinc factors in the denominator are necessary corrections to account for the finite temporal width, $T_\text{flash}$, of the X-ray flare \citep{Epitropakis16}. 

The cross covariance, $h_{AB}$, is a complex number and its argument defines the phase lag of band B with respect to band A. Equation \ref{eq:crosspectrum} shows that at a given frequency $\nu$, this phase lag is equal to the difference of the argument of the transfer function $\Gamma_B$ minus the argument of the transfer function of the reference waveband, $\Gamma_A$, at the same frequency. Thus, we estimate the model phase lag of waveband B with respect to waveband A as:

\begin{equation}
\label{eq:phaselag_AB}
    \phi_{AB} [\nu_j] =  \mathrm{arg}(\Gamma_B[\nu_j])  - \mathrm{arg}(\Gamma_A[\nu_j]) .
\end{equation}

\noindent Meanwhile, for each frequency \textbf{$\nu_j$}, the corresponding time lag may be estimated as:

\begin{equation}
\label{eq:timelag_AB}
    t_{AB} [\nu_j] = \frac{\phi_{AB} [\nu_j]}{2 \pi \nu_j} .
\end{equation}

By convention, a positive lag indicates that variability in the B waveband is delayed with respect to the A band variability. Moreover, the rest of the paper assumes that the lag is always estimated with respect to a bluer waveband, as is customary in observational studies, i.e. $\lambda_A < \lambda_B$ in the above notation. Unless otherwise noted, we use a waveband centered at 1158 \AA~as the reference band throughout the paper.

\begin{figure}[t]
\includegraphics[width=0.95\linewidth,height=0.95\linewidth, trim={0 0 0 0}, clip]{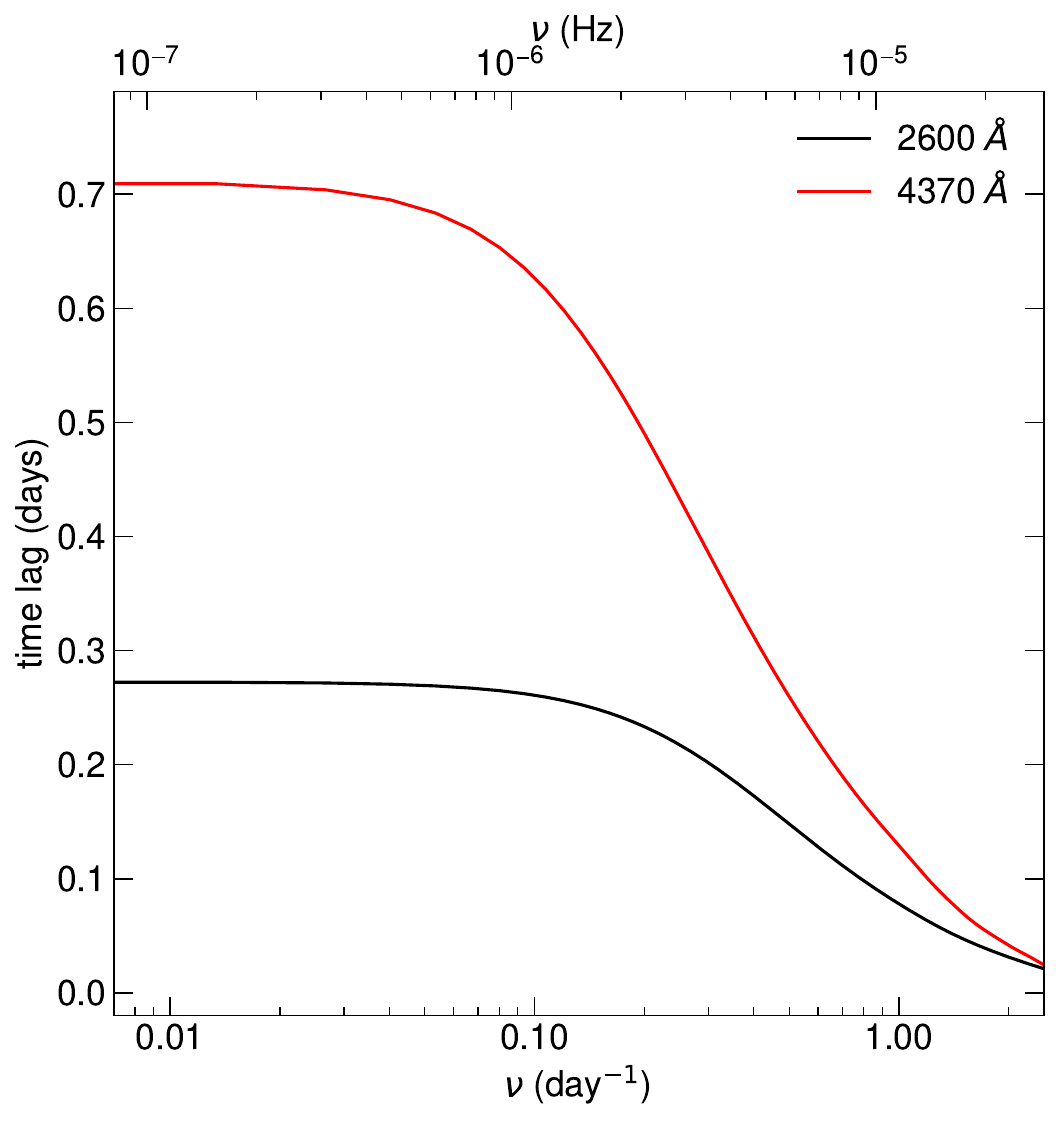}
\caption{Frequency-resolved time lags for two wavebands, centered at 2600 (black) and 4370 $\AA$ (red curve), respectively. The lag is estimated with respect to a reference band centered at 1158 \AA. The fiducial values of Table \ref{tab:param_grid} were assumed in computing the plotted lags. At low frequencies, the lag takes its maximum value forming a plateau, while the time lag decreases at frequencies above a characteristic frequency $\nu_d$, which depends on the effective size of the emitting region in a given waveband. }  
\label{fig:phaselags_example}
\end{figure}

An example of the predicted time lags as a function of Fourier frequency for two wavebands is shown in Fig. \ref{fig:phaselags_example}. At low frequencies, the time lag curve is rather flat and equal to its maximum value. On those long time scales, enough time has passed so that the whole accretion disk has responded to any induced variability. Thus, the time delay between the variability of any two bands is expected to be constant at low frequencies. The exact value of this delay depends on the chosen waveband, with redder wavebands featuring larger lags.

At higher frequencies, or equivalently shorter time scales, the time lag starts to decrease towards a value close to zero. The exact frequency, say $\nu_d$, at which this decrease starts, corresponds roughly to the time needed for the X-ray induced variability to propagate through the entire disk area that is effectively emitting in the considered waveband. Since emission in shorter wavelengths is effectively emitted by more inner parts of the accretion disk, corresponding to emitting areas of smaller spatial extent, $\nu_d$ is expected to increase for smaller wavelengths. This behavior is depicted in Fig. \ref{fig:phaselags_example}, where the lag at $2600$ \AA~starts decreasing from its maximum value at considerably higher frequencies than the $4370$ \AA~lag. 

At frequencies above $\nu_d$, time lags decrease with frequencies. This is due to the fact that X-ray variations are still propagating through the disk on those time scales, and thus a lower lag will be measured. At even higher frequencies, the time lag alters sign due to phase wrapping, similar to the case of X-ray reverberation mapping \citep[e.g.][]{Epitropakis2016b}. As this occurs at frequencies that are not probed by typical contemporary campaigns, we postpone a detailed investigation of this effect to a future work.

%%%%%%%%%%%%%%%%%%%%%%%%%% Table with parameter grid %%%%%%%%%%%%%%%%%%%%%%%%
\begin{table}[]
    % \centering
    \caption{Parameter grid over which we calculate the predicted frequency resolved time lags}
    \begin{tabular}{lr}
    \hline
    Parameter       &  values   \\     
    \hline
    
    $M_\text{BH}$ (10$^8 M_\odot$)      &  0.01, 0.05, \textbf{0.1}, 0.5, 1, 5    \\
    % $h_{X}$  ($R_g$\tablenotemark{a})      & 2.5, 5, \textbf{10}, 20,  40,  60,  80, 100          \\
    $h_{X}$  ($R_g$)      & 5, \textbf{10}, 20,  40,  60,  80, 100          \\
    $\dot{m}_\text{Edd}$      &  0.005, 0.01, 0.02, \textbf{0.05}, 0.1, 0.2, 0.5, 0.8     \\
    $L_\mathrm{trans}$      &  0.01, 0.02, 0.05, \textbf{0.1}, 0.2, 0.5,  0.8      \\
    % $\Gamma_{X}$      &  1.70     \\
    $\theta$ ($^\circ$)      &  5, 20, \textbf{40}, 60, 80      \\
    \tableline 
    \end{tabular}
    \label{tab:param_grid}
    \tablecomments{Numbers in bold denote the fiducial values (see Sect. \ref{sec:depend}). The corona height is measured in gravitational radii, the accretion rate is in units of the Eddington accretion limit and the corona luminosity, $L_\mathrm{trans}$, is given as the fraction of the total accretion power.}
    % \tablenotetext{a}{$R_g$ denotes the gravitational radius.}
\end{table}

%%%%%%%%%%%%%%%%%%%%%%%%%%%%%%%%%%%%%%%%%%%%%%%%%%%%%%%%%%%%%%%%%%%%%%%%%%%%

\section{Dependence on physical parameters}
\label{sec:depend}

In this Section, we discuss how the shape of the time lags as a function of frequency varies for different system configurations. Our discussion focuses mostly on the dependence of the lag amplitude, that is the constant lag at low frequencies, and the dependence of the characteristic frequency $\nu_d$ on the various model parameters. As already mentioned, both of these depend primarily on the effective size of the emitting region.

\begin{figure*}
\includegraphics[width=0.32\linewidth,height=0.37\linewidth, trim={0 0 0 0}, clip]{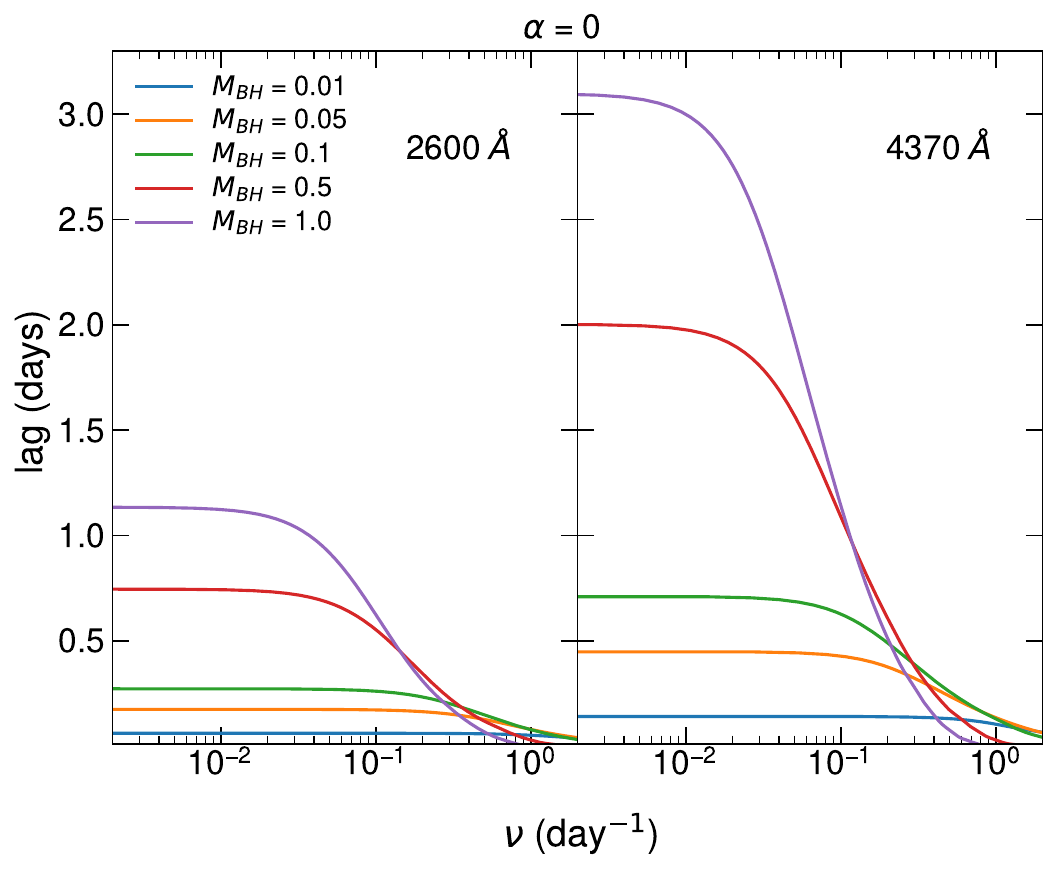}
\includegraphics[width=0.32\linewidth,height=0.37\linewidth, trim={0 0 0 0}, clip]{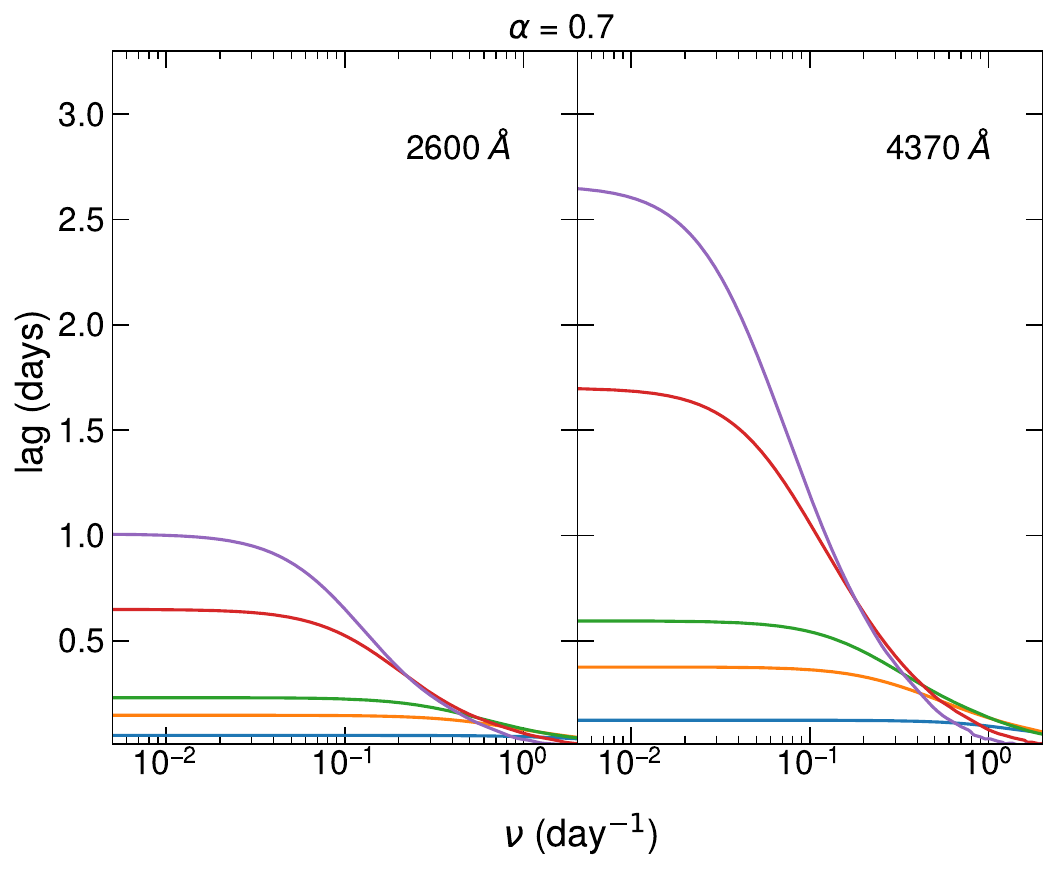}
\includegraphics[width=0.32\linewidth,height=0.37\linewidth, trim={0 0 0 0}, clip]{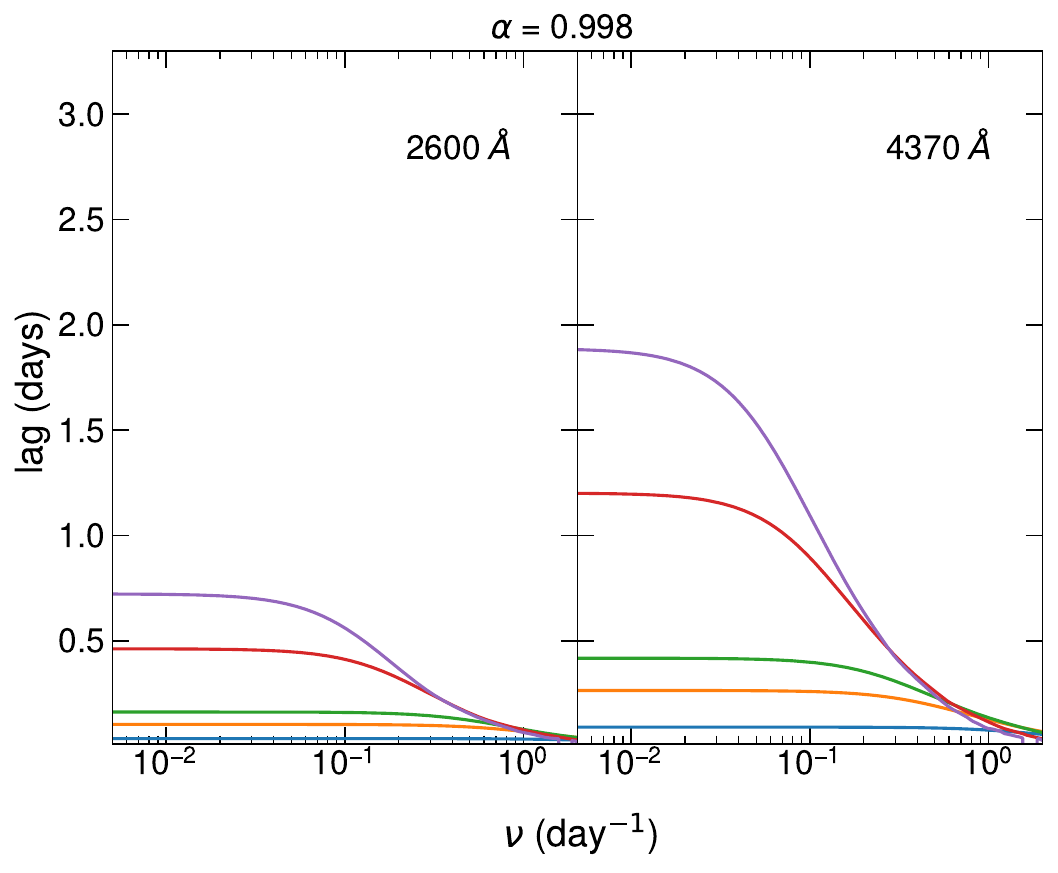}
\caption{The predicted frequency-resolved time lags for spin 0 (left), 0.7 (middle), and 0.998 (right), for different black hole masses, at two wavebands (2600 and 4370 \AA). The time lags are computed using the simulated response functions and eq. \ref{eq:timelag_AB}. The values of black hole mass in the inset are in units of 10$^8 M_\odot$.  }    
\label{fig:phaselags_mass}
\end{figure*}

In order to explore how the expected time lags alter with the various physical properties, we computed the disk response function for different wavebands and for a range of corona heights, $h_X$, black hole masses, $M_\text{BH}$, accretion rates, $\dot{m}_\text{Edd}$, inclination, $\theta$, and power released in the X-ray corona, $L_\text{trans}$\footnote{In the KYNXiltr model, the X-ray corona power may be due to (an unknown mechanism of) external heating or directly obtained from the accretion power. In the latter case, the accretion power released within a specific radius is transferred to the corona \citep{Dovciak2022}. We use the latter as our fiducial case, but we explore both options, when we study the dependence of the phase lags on X-ray luminosity.}. The range of values considered for each parameter are listed in Table \ref{tab:param_grid}. In exploring the dependence on a single parameter, we keep the rest of them fixed at their fiducial value, which is denoted in bold in Table \ref{tab:param_grid}. We have also assumed that the disk has an outer radius of $R_{out} = 10000 R_g$ and a color correction factor of $f_{col}=2.4$. The energy spectrum of the incident X-ray emission is assumed to follow a power law with photon index 2 and high energy cutoff $E_{cut} = 300 ~\text{keV}$. 

We discuss our main findings below, while most of the plots are shown in Appendix \ref{sec:plots}. For simplicity, we show the results for only two wavebands centered at 2600 \AA~ and 4370 \AA, which resemble the commonly used UVW1 and B wavebands \citep[e.g., Table 6 of][]{Fausnaugh2016}, and for three values of the black hole spin, $\alpha = 0, 0.7,$ and $0.998$. 

\subsection{Dependence on $M_\text{BH}$, $\alpha$, $h_X$, and $\dot{m}$}

Figure \ref{fig:phaselags_mass} displays the frequency-resolved time lags for various black hole masses. As the mass of the black hole increases, the lag amplitude also rises, while the frequency $\nu_d$ decreases. This behavior can be explained by considering how the physical size of the accretion disk changes with black hole mass. For a larger black hole, emission at a given wavelength comes from a more spatially extensive region of the accretion disk. Consequently, it takes longer for the induced variations to propagate through this disk. This leads to a decrease in $\nu_d$ and an increase in time lags at lower frequencies.

The same figure highlights how the time lags depend on the black hole spin and the considered waveband (see also Figs. \ref{fig:phaselags_height}-\ref{fig:phaselags_lumin_plus}). The emission from redder wavebands originate from a larger region, as outer parts of the accretion disk contribute significantly at these wavelengths. Therefore, the lags increase with the wavelength, similarly to the case of time domain estimated lags \citep{Kammoun2021}. For the same reason, the frequency $\nu_d$ is shifted towards smaller values as we consider variability in longer wavelengths.

The time lags depend on the black hole spin as well. At a given wavelength, the time lag amplitude decreases and $\nu_d$ increases with increasing black hole spin.
This is because the spin value effectively sets the accretion rate in physical units, as we keep $\dot{m}$ fixed to a fraction of the Eddington limit. Higher spin values correspond to larger radiative efficiencies, requiring thus a smaller amount of mass to be accreted in order to match the selected $\dot{m}$. Therefore, the disk temperature is lower and the emitting region at a specific waveband is less extended for higher values of the spin.

\begin{figure*}[]
\includegraphics[width=0.32\linewidth,height=0.37\linewidth, trim={0 0 0 0}, clip]{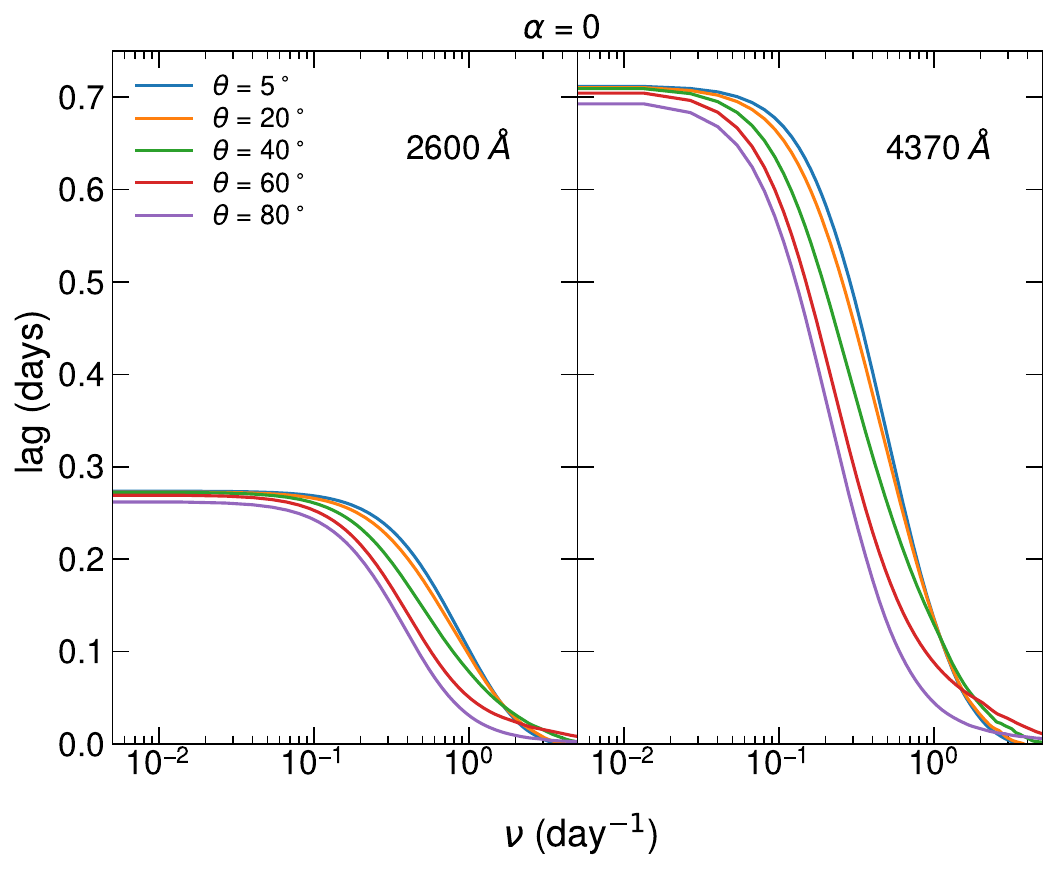}
\includegraphics[width=0.32\linewidth,height=0.37\linewidth, trim={0 0 0 0}, clip]{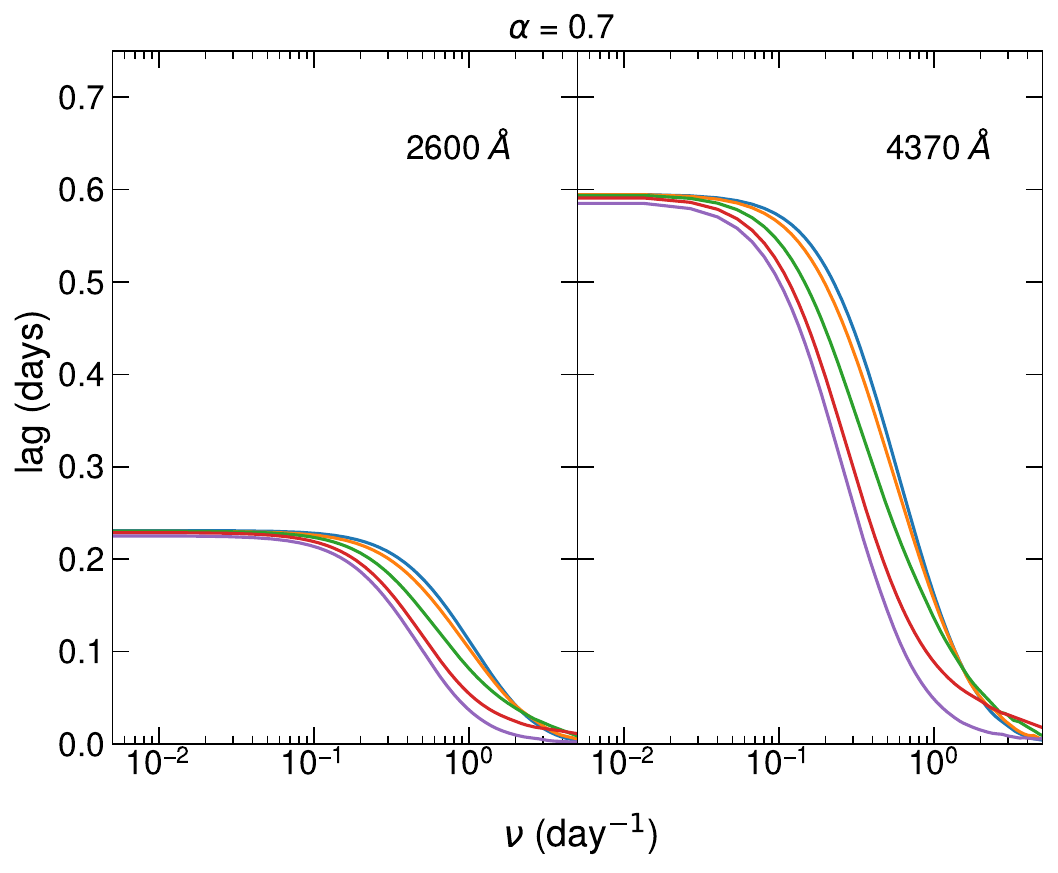}
\includegraphics[width=0.32\linewidth,height=0.37\linewidth, trim={0 0 0 0}, clip]{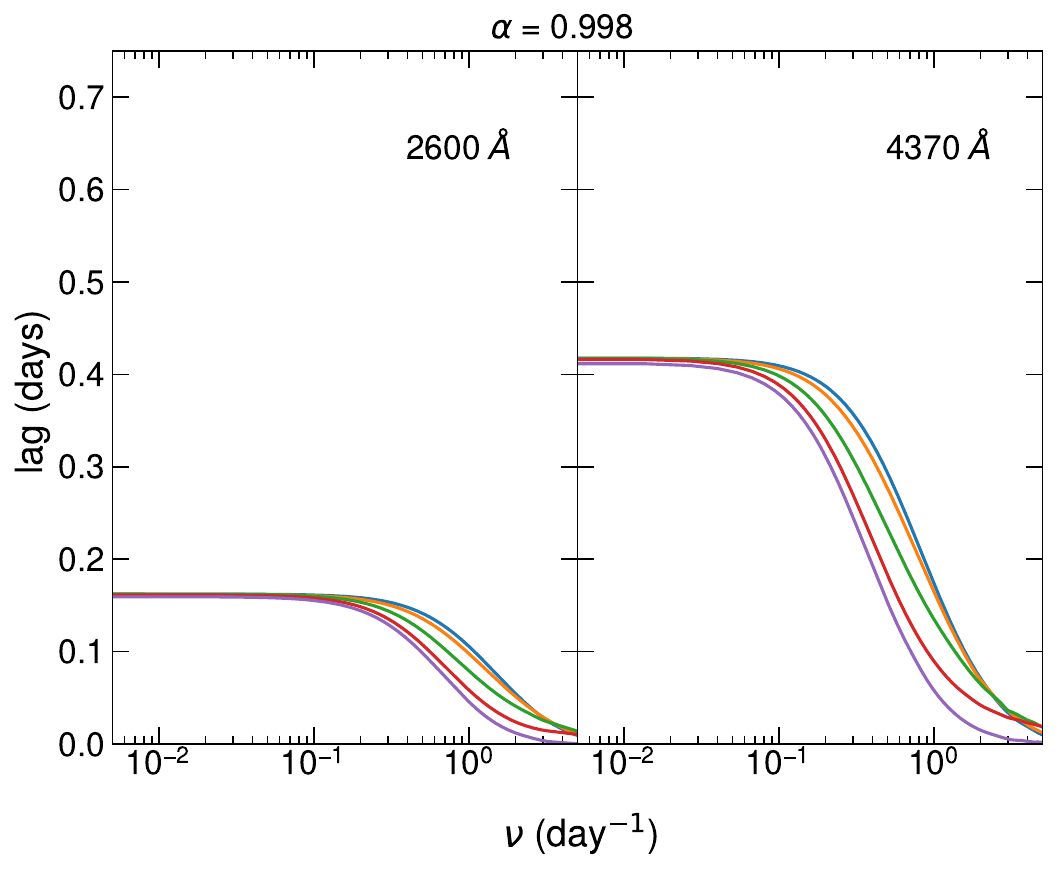}
\caption{Same as Fig. \ref{fig:phaselags_mass} for different values of the source inclination.}    
\label{fig:phaselags_incl}
\end{figure*}

Further, the predicted time lags feature a more subtle dependence on the height of the X-ray source. An increase in height results in a larger flux of incident X-rays onto the accretion disk, subsequently raising the disk temperature. As a result, one would expect the time lags to increase with the height, while the frequency $\nu_d$ decreases, as is illustrated in Fig. \ref{fig:phaselags_height}.

Finally, the frequency-resolved lags feature a strong dependence on the accretion rate, as is expected. A larger accretion rate increases the temperature of the accretion disk, leading to a larger effective size for the emitting region at a specific waveband. Hence, the low-frequency lag shall increase with the accretion rate, while $\nu_d$ decreases. These trends are evident in Fig. \ref{fig:phaselags_mdot}.

\subsection{Dependence on $L_\mathrm{trans}$}

The model used allows for the X-ray source to be heated either externally or through the accretion process. In the external heating scenario, the mechanism powering the X-ray source is assumed to have no impact on the accretion disk, which initially follows the typical \cite{NT73} prescription up to its innermost stable circular orbit. Conversely, when heating is driven by the accretion process, the energy released in the X-ray corona is extracted from the inner accretion disk. In this case, the disk is initially cold within a specific radius, though it still extends to the innermost circular orbit \citep[see][for a detailed discussion of this effect]{Dovciak2022}. We opted to investigate both scenarios.

Figure \ref{fig:phaselags_lumin_minus} presents the results for an externally heated corona. As more power is released in the corona, its flux increases, leading to more X-rays being absorbed by the accretion disk. This raises the disk temperature and thus, emission at a given waveband originates from a more spatially extended area. Consequently, we observe a larger lag amplitude and smaller $\nu_d$ for higher values of $L_\text{trans}$.

The situation is more complex when the X-ray power is transferred from the accretion disk. Then, the effect of an inner accretion disk lacking internal heating might manifest in the predicted lags. As discussed in detail by \cite{Kammoun2023}, this effect modifies considerably the disk response function at early times, when the observed flux is dominated by the innermost parts of the disk. In the context of frequency-resolved lags, Fig. \ref{fig:phaselags_lumin_plus} shows that the overall picture remains qualitatively similar to that of an externally heated corona, with increased corona power resulting in larger lags and reduced $\nu_d$. However, this trend does not hold for low black hole spin values combined with high $L_\text{trans}$. In this extreme case, a substantial area of the inner disk (of the order of 100 $R_\text{g}$) lacks internal heating and the main source of power for the disk is now X-ray illumination. As a result, and since roughly half of the X-rays are emitted away from the disk, the emission at a given waveband will originate from a smaller area, leading to a reduced lag amplitude and a higher $\nu_d$.

\subsection{Dependence on inclination}

Perhaps, the most interesting evolution of phase lags is found when one considers their dependence on inclination (Fig. \ref{fig:phaselags_incl}). The lag amplitude shows minimal variation with inclination angle, whereas the frequency $\nu_d$ decreases as the inclination increases. While the disk response does depend on inclination, its centroid remains rather unchanged for different inclination values; hence the time lags measured via cross correlation techniques do not depend significantly on inclination. Similarly, the low-frequency amplitude of the frequency-resolved time lags plotted in Fig. \ref{fig:phaselags_incl} remains more or less constant, independent of inclination. But this is not the case with the time lags dependence on frequency, which features a clear dependence on inclination. Obviously, the differences on the shape of the response function are captured by the time-lags functions. 
Notably, this strong dependence suggests that frequency-resolved time lags could potentially be utilized to constrain the source inclination angle,  offering a novel approach to estimating this property that is very challenging to determine via other methods.

\section{Comparison to observational data}
\label{sec:5548}

\begin{figure*}[t]
\includegraphics[width=\linewidth,height=0.4\linewidth, trim={70 0 70 0}, clip]{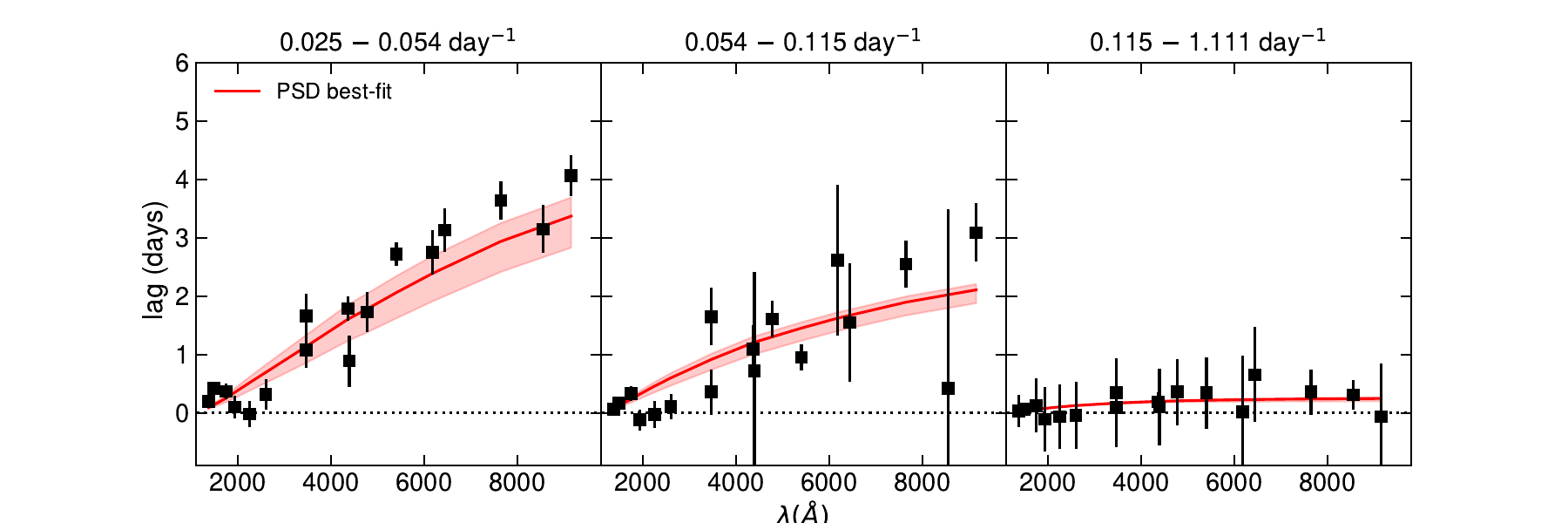}
\caption{Frequency-resolved time lag spectra of NGC 5548 in three frequency bins, measured with respect to the 1158~\AA~ light curve (black squares; values retrieved from \cite{Cackett2022}). The red lines show the predicted lag spectra using the best-fit model of \cite{Panagiotou2022ApJ}. The shaded regions denote the 3-$\sigma$ confidence area. }    
\label{fig:phaselags_5548}

\end{figure*}

In order to test the results of our theoretical study against observational measurements, we compare our predictions to the case of NGC 5548. This source was the target of one of the earliest broadband reverberation mapping campaigns, during which it was observed with subdaily cadence for a few months across a range of wavebands \citep[STORM 1,][]{DeRosa2015}.
\cite{Cackett2022} used the corresponding light curves to measure the frequency-resolved time lags of NGC 5548 for every UV and optical waveband following the maximum likelihood approach of \cite{Zoghbi2013}. 

These authors estimated the time lag spectrum in 4 frequency bins, logarithmically  spaced from 0.012 to 1.111 day$^{-1}$. Here, we consider the results for three of these bins and postpone the analysis of the lowest frequency bin to the subsequent section.

We used the best-fit results of \cite{Panagiotou2022ApJ} to estimate the expected frequency-resolved time lags when these are solely due to the X-ray illumination of the disk. In that work, we had shown for the first time that the X-ray thermal reverberation of the accretion disk can explain well both the UV/optical variability power spectra and the CCF time lags. In fact, by fitting simultaneously the observed PSDs and CCF time lags, we were able to constrain the X-ray corona's height and the accretion rate to $h_\text{X} = 30 \pm 5~\text{R}_\text{g}$ and $\dot{m}_\text{Edd} = 0.06 \pm 0.02$, respectively. 

We calculate the predicted time lags in each frequency bin using these best-fit values. In this way, we will be able to test if the X-ray disk reprocessing can explain both the PSDs and the frequency-resolved time lags at different wavebands in a self-consistent way. Following \cite{Panagiotou2022ApJ}, we assume an externally heated X-ray corona emitting a power law spectrum with photon index $\Gamma = 1.5$, high energy cutoff $E_\text{cut} = 300~$keV, and an observed X-ray luminosity of $L_\text{2-10keV} = 0.0034~L_\text{Edd}$, as found by \cite{Mathur2017}. We further assume a black hole mass of $M_\text{BH} = 5.2 \cdot 10^7 M_\odot$, obtained from the AGN black hole mass database \citep{Bentz2015} for  a virial factor f=4.3 \citep{Grier2013}, an inclination angle of $\theta = 40 ^\circ$ \citep{Pancoast2014}, and a highly spinning black hole, $\alpha=1$, which is preferred by the broadband spectral analysis of this source \citep{Dovciak2022}. 

Figure \ref{fig:phaselags_5548} presents the predicted time lags as a function of wavelength for the three considered frequency bins, alongside the measured time lags reported by \cite{Cackett2022}. The plot reveals a very good agreement between the model predictions and observations within the statistical uncertainties. Notably, the predicted curve closely matches the observed lags even in the lowest frequency bin, indicating no evidence for an additional component down to temporal frequency of 0.025 day$^{-1}$. Interestingly, the shape of the model lags seems to differ from the commonly used $\tau \propto \lambda ^{4/3}$ relation, as not all regions of the accretion disk have responded to the same degree to the X-ray illumination on those time scales.

We emphasize that no fit has been applied to the plotted data. The model curves are derived solely from the source configuration that accurately reproduced the measured PSDs. The observed agreement becomes even more remarkable, if we consider that important parameters, such as the black hole mass and the source inclination, have been kept fixed in the PSD fitting and in predicting the time lags. Still, the predicted curves follow closely the measurements, with no systematic features of disagreement. It is also worth mentioning that wavebands above 6000 \AA~were not considered in the PSD fit, which might explain the slight underestimation of the time lags in the i and z bands; while at these long wavelengths the choice of the outer radius of the disk and the model assumption of a razor thin accretion disk become important. We conclude that the X-ray illumination of the accretion disk effectively accounts for both the UV/optical power spectra and observed time lags across different temporal frequencies in a self-consistent way.

\subsection{Observed lags below 0.025 day$^{-1}$}

\begin{figure}[]
\includegraphics[width=\linewidth, trim={0 0 20 0}, clip]{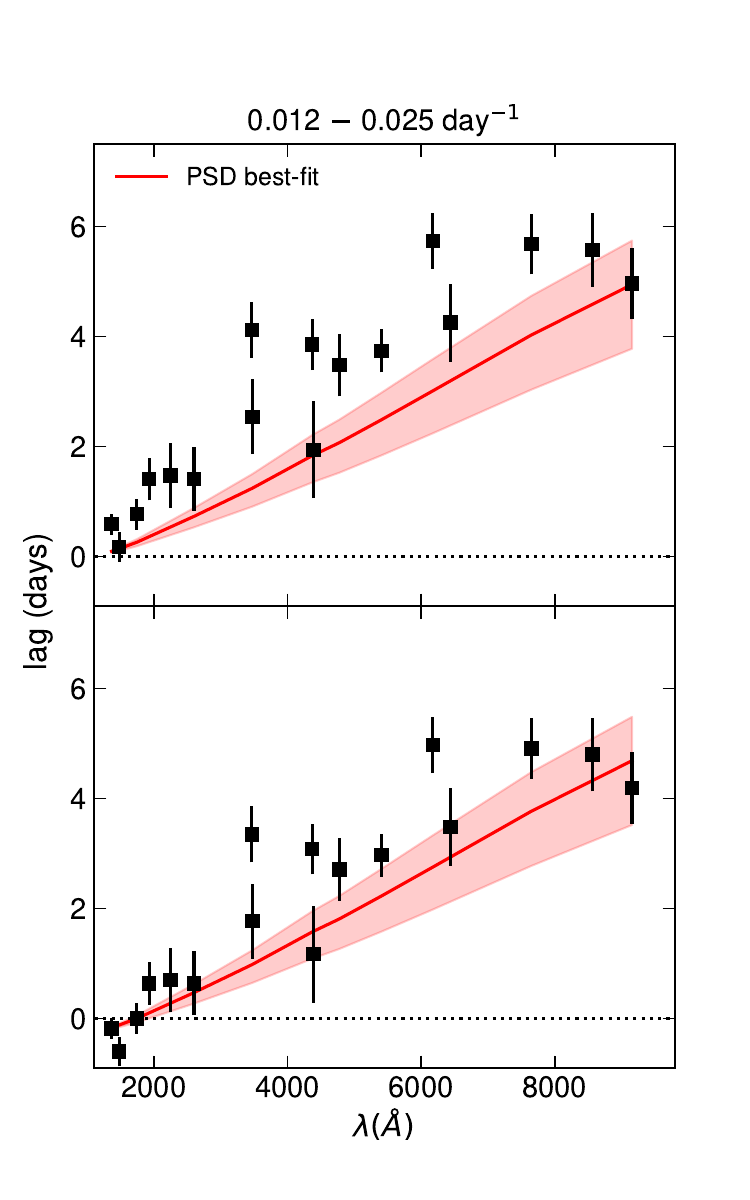}
\caption{Frequency-resolved time lag spectra of NGC 5548 in the lowest frequency bin considered by \cite{Cackett2022}). Similarly to Fig. \ref{fig:phaselags_5548}, the red lines denotes the predicted lag spectra using the best-fit model of \cite{Panagiotou2022ApJ}, with the shaded regions indicating the 3-$\sigma$ confidence area. The upper panel shows the time lags with reference to the 1158 \AA~emission, while in the lower panel the emission at 1746 \AA~has been used as the reference band.}
\label{fig:phaselags_5548_lowfreq}

\end{figure}

In addition to the time lags considered above, \cite{Cackett2022} estimated the frequency-resolved time lags of NGC 5548 in an even lower frequency bin, from 0.012-0.025 day$^{-1}$. The corresponding values are plotted in the upper panel of Fig. \ref{fig:phaselags_5548_lowfreq}, along with the model predictions estimated as above. The model predictions seem to systematically underestimate the observed lags in this case. However, caution needs to be exercised when interpreting the results at these low frequencies. This frequency bin spans only three Fourier frequencies, close to the minimum frequency probed by the light curves, and hence, the measured lag may be of higher uncertainty. This is especially important as such uncertainties may affect the lag of the reference band.

To demonstrate that, the lower panel of Fig. \ref{fig:phaselags_5548_lowfreq} plots the same time lags when measured with respect to the 1746 \AA~emission. In this case, a better agreement is found between the observed data and the model, with the lags in most wavebands being consistent with the predictions within statistical uncertainties. 

This significant difference suggests that the current data set is not sufficient to conclude with high confidence if the X-ray reverberation can account for the observed lags on these long time scales. Longer light curves are necessary to reach such a conclusion, as they would allow us to constrain better the corresponding lags. While these are the time scales where the BLR contribution is expected to be more important, it should also be mentioned that other variability mechanisms may contribute significantly in these frequencies, as has been demonstrated in the case of the nearby AGN Fairall 9, which was observed for a longer period \citep[e.g.][]{Jua2020Fairall9, Yao2023, Partington2024}. Such variability mechanisms may originate in the disk and could, for example, include distributed regions of varying temperature across the disk \citep[e.g.][]{Dexter2011} or mass fluctuations propagating through the accretion disk \citep[][]{Lyubarskii1997}.

\section{Conclusions} 
\label{sec:conclude}

The analysis of the frequency-resolved lags in variable sources provides an elegant and powerful means to differentiate among the various mechanisms driving variability across different time scales. This method has recently been applied to study the UV/optical timing properties of AGN \citep{Cackett2022, Lewin2023, Lewin2024}.
In this work, we have investigated the frequency-resolved lags of AGN expected when their UV/optical variability is driven by the X-ray illumination of the accretion disk.

We present the predicted time lags as a function of frequency for a range of physical parameters. The overall shape of these predicted lags aligns closely with those derived from observational data. Specifically, the lag reaches its maximum value at low frequencies, typically at less than 0.05 day$^{-1}$ for the optical wavebands and a source with $M_\text{BH} = 5 \cdot 10^{7} M_\odot$ (Fig. \ref{fig:phaselags_mass}), while it exhibits a sharp decline towards values near zero at frequencies higher than a bending frequency $\nu_d$. 

We have also explored the dependence of frequency-resolved lags on the various physical properties of AGN (Section \ref{sec:depend}). 
The amplitude of the time lags at low frequencies increases with the black hole mass, wavelength, corona height, X-ray luminosity and accretion rate. Conversely, it decreases as the black hole spin increases, while it remains rather unchanged for different values of the inclination angle. In contrast, the characteristic frequency $\nu_d$ depends strongly on inclination, featuring lower values in more inclined systems. The frequency $\nu_d$ was also found to increase with the black hole spin, while it decreases with increasing black hole mass, accretion rate, wavelength, corona height and X-ray luminosity. These predictions can be compared directly with observational results, as the timing properties of more sources are analyzed in the Fourier space. 

It should be noted that not all parameters affect the predicted lags to the same degree. For example, the X-ray luminosity and corona height seem to have a subtler impact on the expected lag values, suggesting it would be hard to constrain these parameters with high precision when fitting observed time lags, unless the corresponding data are of very high quality. However, the power spectra are predicted to feature a strong dependence on these parameters \citep{Panagiotou2022ApJ}. Thus, the combined modeling of power spectra and time lags offers a powerful approach to obtain robust constraints of the various parameters. In addition, other physical properties, such as the accretion rate and the disk inclination, are predicted to have a stronger impact on the expected lags, which underscores the potential of frequency-resolved lags to constrain critical parameters of AGN that are otherwise challenging to measure.

Furthermore, we compare our model predictions against the observational measurements of the nearby Seyfert galaxy NGC 5548, using the best-fit PSD model parameters to estimate the predicted time lags (Sect. \ref{sec:5548}). The observed lag spectra are in good agreement with the model. Thus, the same model configuration reproduces well the observed power spectra and frequency-resolved time lags in a self-consistent manner. To the best of our knowledge, this is the first and only physical model that has successfully achieved this concordance. We infer that the X-ray illumination of the accretion disk accounts for the complete frequency-resolved variability properties of this AGN.

It is important to stress out that the lowest frequency bin plotted in Fig. \ref{fig:phaselags_5548} spans from 0.025 to 0.054 day$^{-1}$, corresponding to time scales of $18-40$ days. This is significantly longer than the H$\beta$ lag for this source, typically 2-7 days \citep[e.g.][]{Bentz2009, Pei2017, DeRosa2018}, which is expected to be comparable, or even larger than the time lag of diffuse continuum emission \citep{Netzer2022}. Consequently, any significant contribution to the lags from diffuse continuum emission originating from the BLR should be evident within this frequency range. However, we find no evidence for such an additional variability component at these frequencies. If the diffuse continuum from the BLR does indeed contribute to the observed variability, this contribution shall be minimal, and perhaps becomes important only in specific wavebands.

We also wish to highlight that some caution needs to be exercised when examining time lags at even lower frequencies. For example, \cite{Uttley2003} found that the UV variability of NGC 5548 exceeds its X-ray variability on these time scales, suggesting that mechanisms beyond X-ray reverberation, such as intrinsic disk variations, may dominate the observed variability at these low frequencies.

In our analysis, we simulated the predicted time lags for the model parameters that reproduce the PSDs. Ideally, one would like to fit the two simultaneously. We are currently working on updating the KYNXiltr software and on addressing several computational complexities, with the goal to present a public code that will allow such a fit in the near future. In its current version, the code can be used to produce the response function of the disk for different configurations of the system and in different wavebands. This can already be useful if, for instance, one wishes to explore the range of time lags expected for a source for various heights, black hole spins and perhaps inclination angles, when the X-ray luminosity, the black hole mass and accretion rate are independently known. 

In conclusion, our analysis shows that \uvot\ frequency-resolved time lags due to X-ray illumination of the accretion disk may take a range of values, which, in the case of NGC 5548, closely match the observational measurements. We intend to expand our analysis to more sources with, for example, different accretion rates or light curve properties, in order to further test our main findings in different physical or frequency regimes.

% \begin{acknowledgments}

% \section*{Acknowledgments}

% \end{acknowledgments}

\bibliography{phaselags_predict}{}
\bibliographystyle{aasjournal}

\appendix

\section{Plots of the frequency-resolved time lags}
\label{sec:plots}

In this section, we present plots of the expected time lags vs frequency for all the physical parameters we consider, except for the black hole mass and inclination that are shown in Sect. \ref{sec:depend}. For each configuration, the lags have been estimated with respect to a reference band centered at 1158 \AA.

\begin{figure}[h]
\includegraphics[width=0.32\linewidth,height=0.37\linewidth, trim={0 0 0 0}, clip]{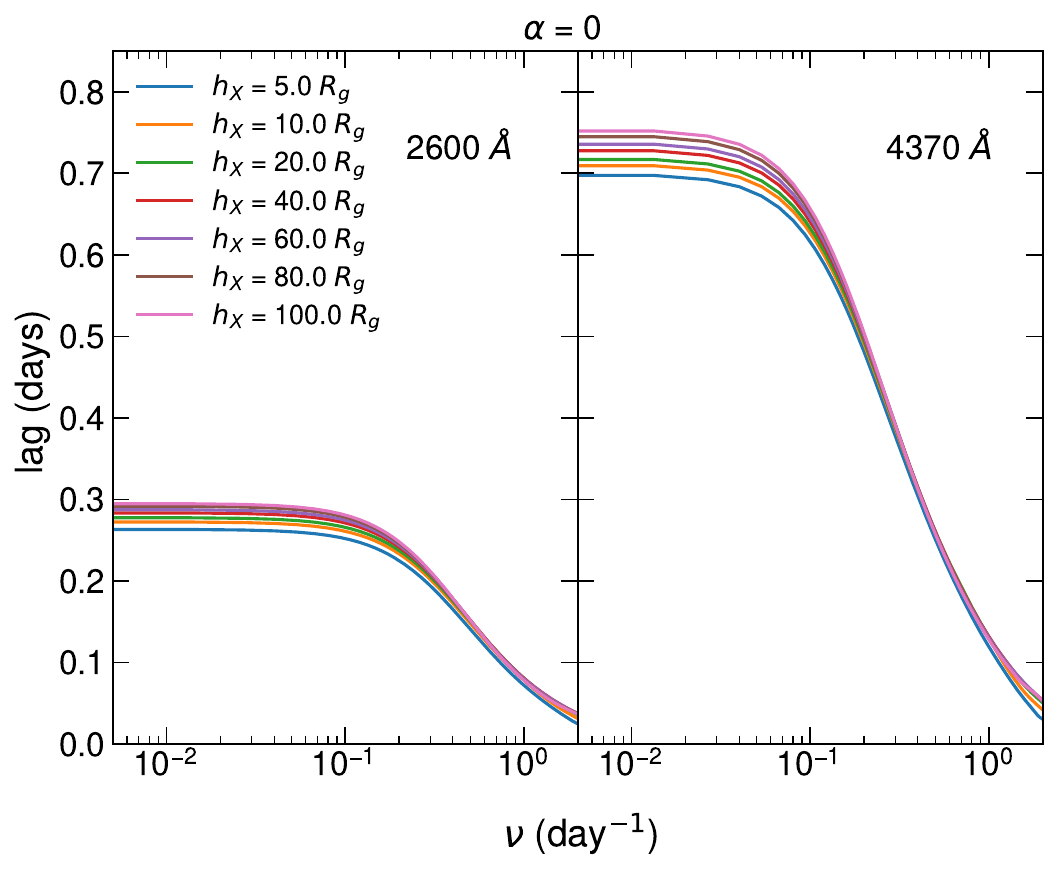}
\includegraphics[width=0.32\linewidth,height=0.37\linewidth, trim={0 0 0 0}, clip]{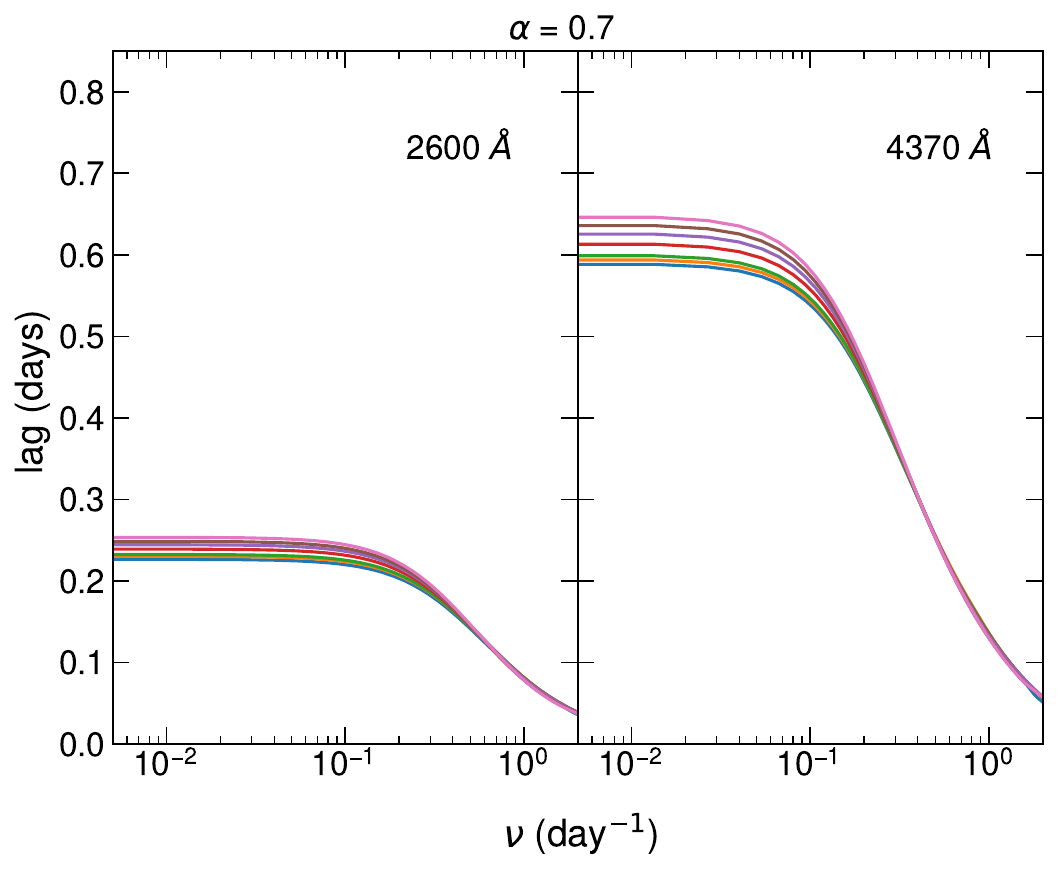}
\includegraphics[width=0.32\linewidth,height=0.37\linewidth, trim={0 0 0 0}, clip]{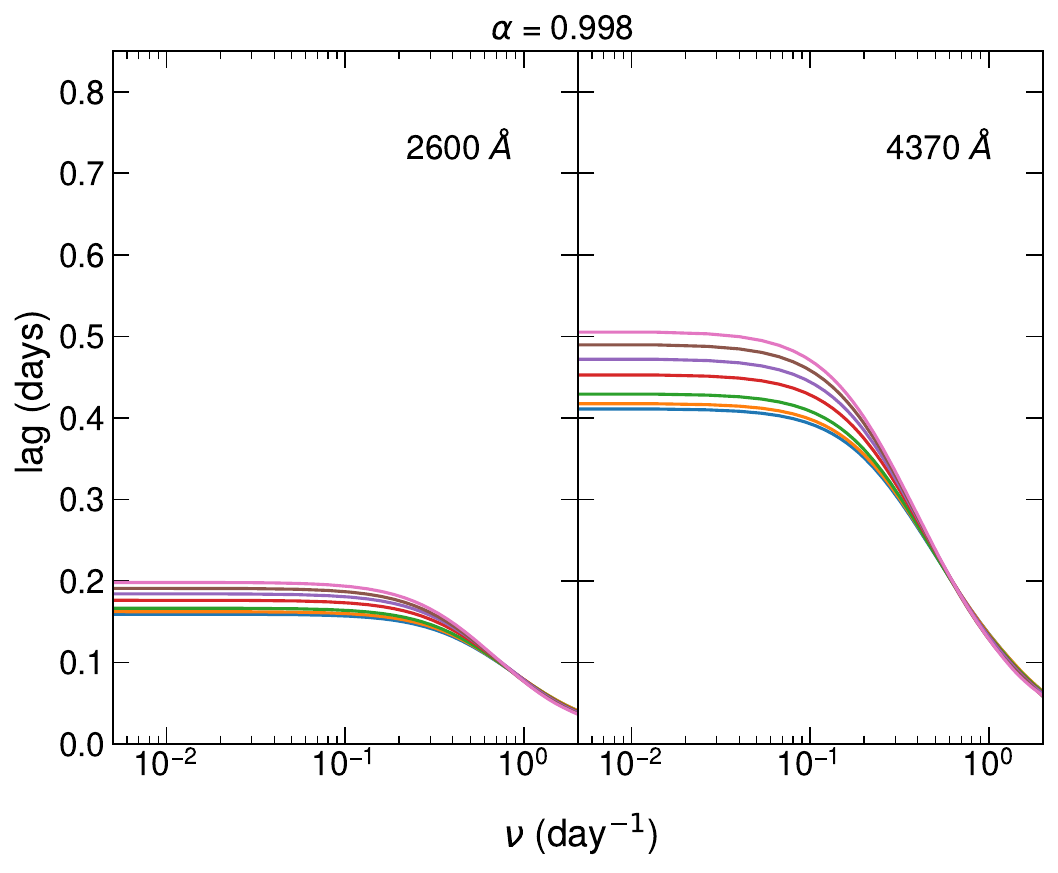}
\caption{The predicted frequency-resolved time lags for spin 0 (left), 0.7 (middle), and 0.998 (right), and at two wavebands (2600 and 4370 \AA), for different heights of the X-ray source, measured in gravitational radii.}    
\label{fig:phaselags_height}
\end{figure}

\begin{figure}[h]
\includegraphics[width=0.32\linewidth,height=0.37\linewidth, trim={0 0 0 0}, clip]{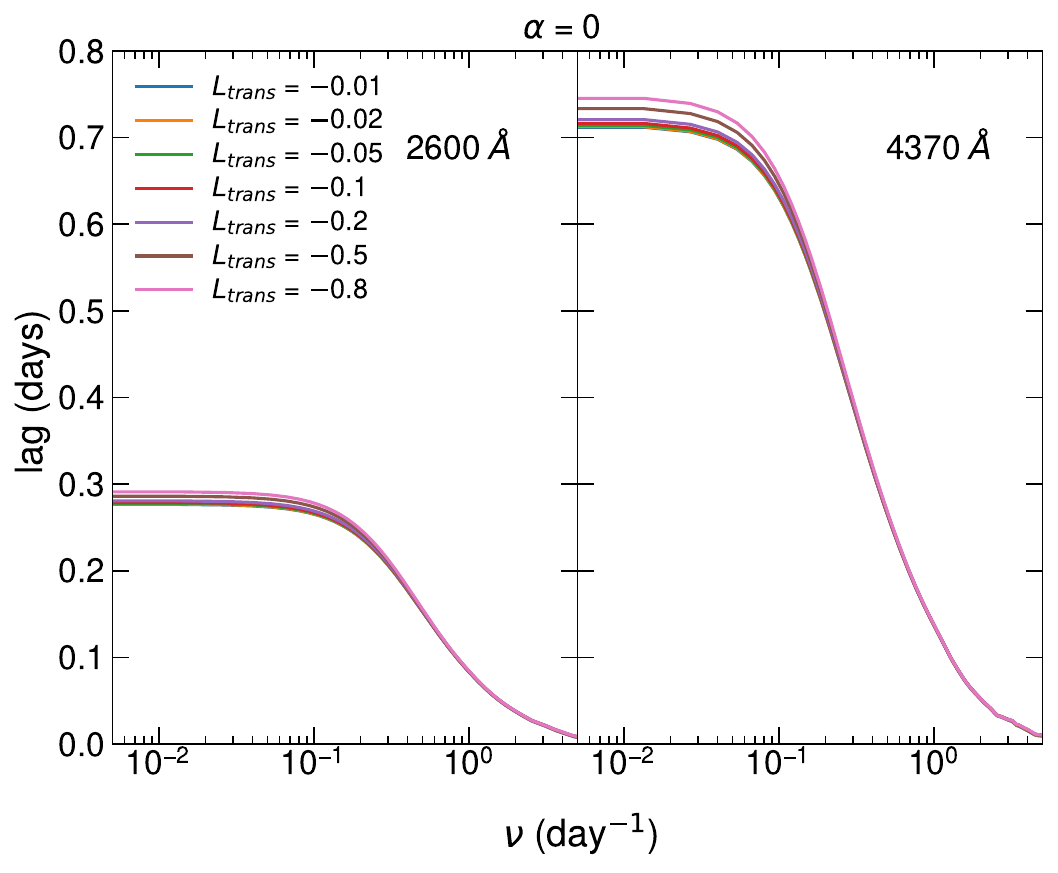}
\includegraphics[width=0.32\linewidth,height=0.37\linewidth, trim={0 0 0 0}, clip]{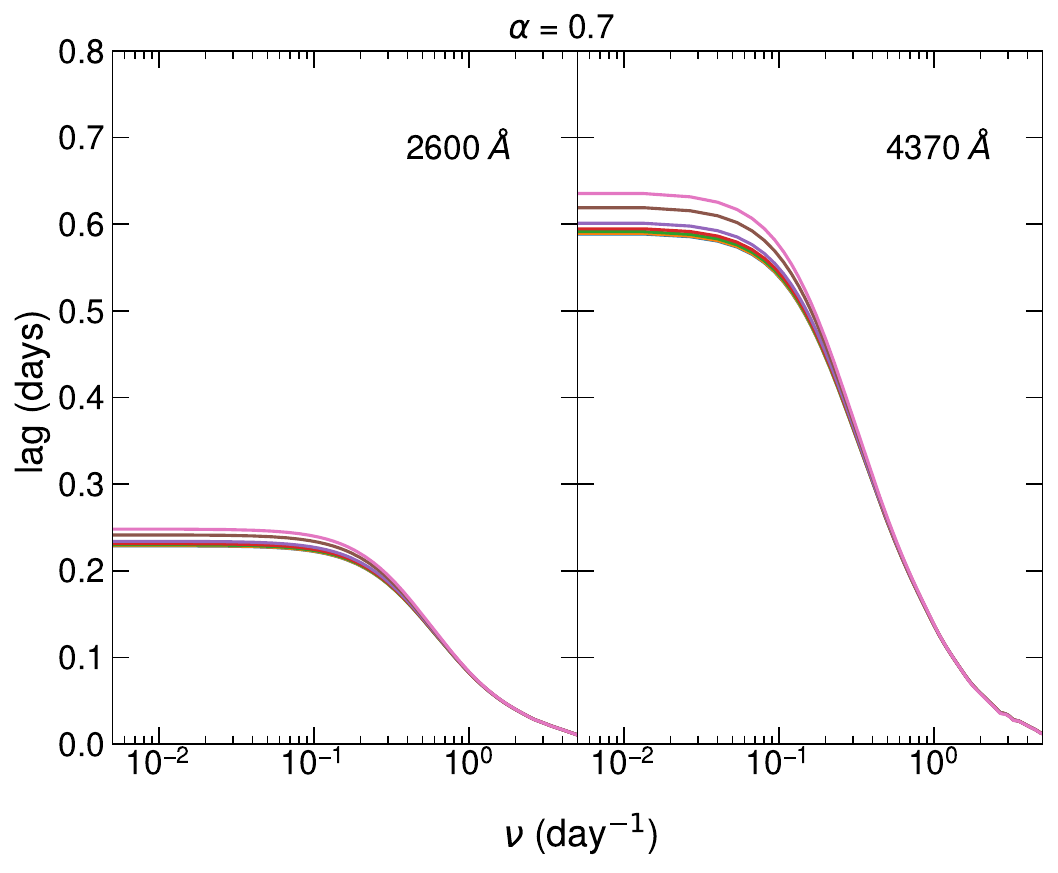}
\includegraphics[width=0.32\linewidth,height=0.37\linewidth, trim={0 0 0 0}, clip]{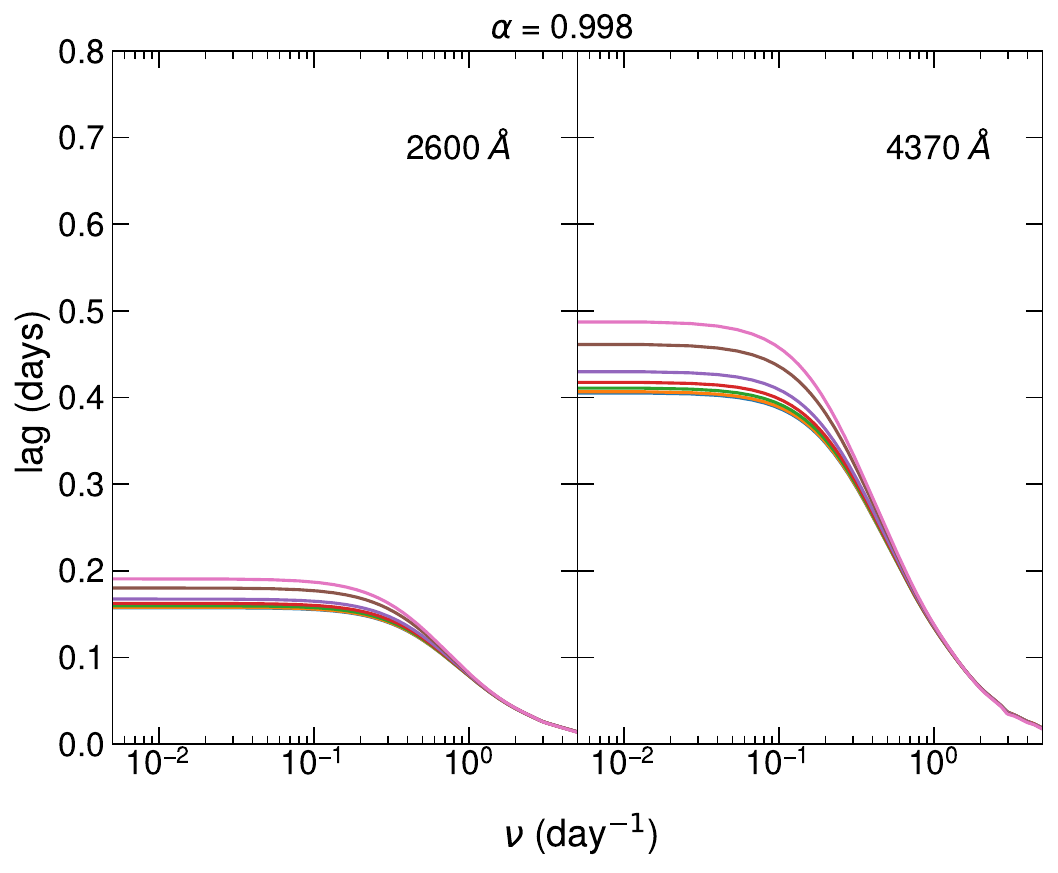}
\caption{Same as in Fig. \ref{fig:phaselags_height} for different values of the X-ray corona power. Here, the X-ray corona is assumed to be heated by an unknown external mechanism. The minus sign in the figure's inset is used to denote that.}    
\label{fig:phaselags_lumin_minus}
\end{figure}

\begin{figure}[h]
\includegraphics[width=0.32\linewidth,height=0.37\linewidth, trim={0 0 0 0}, clip]{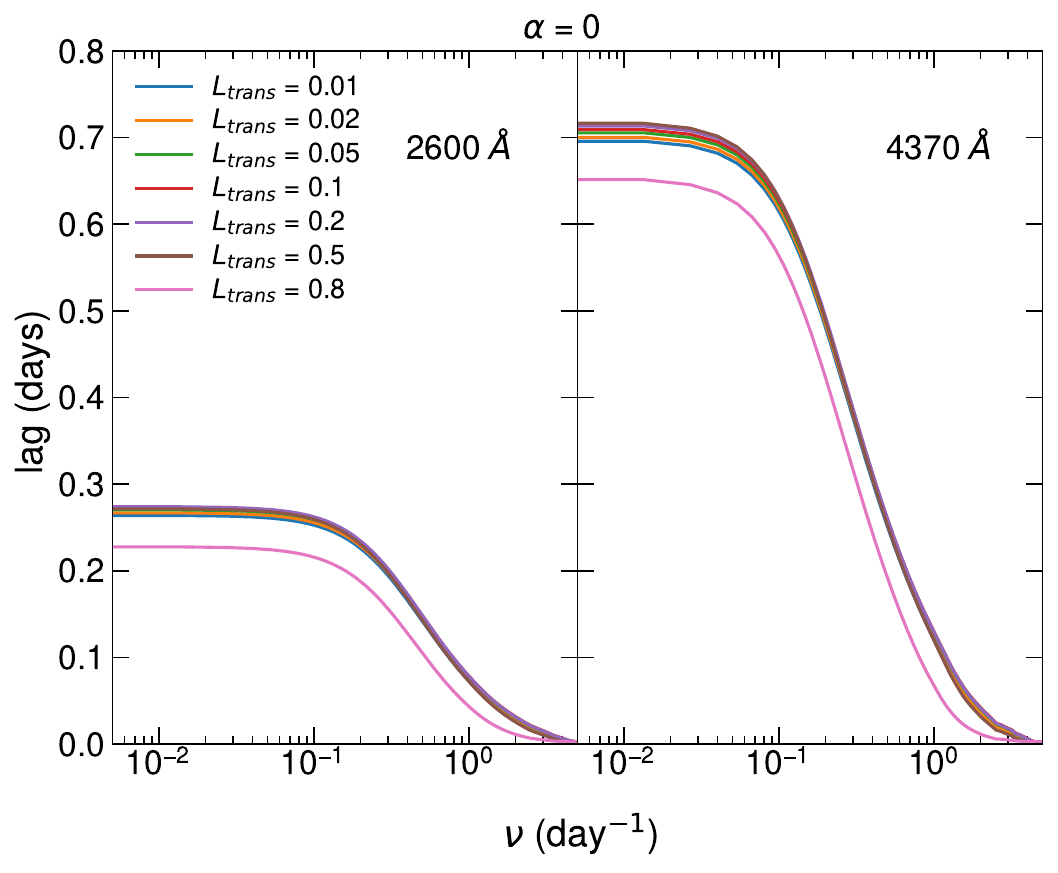}
\includegraphics[width=0.32\linewidth,height=0.37\linewidth, trim={0 0 0 0}, clip]{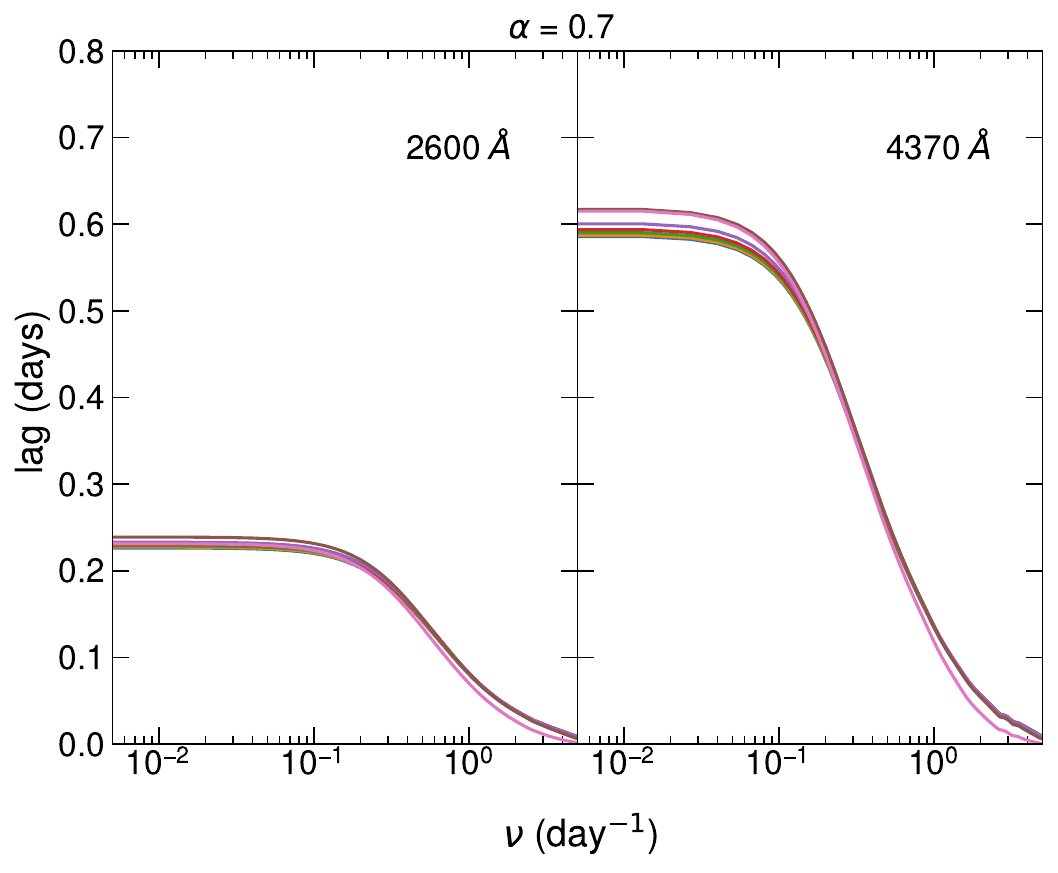}
\includegraphics[width=0.32\linewidth,height=0.37\linewidth, trim={0 0 0 0}, clip]{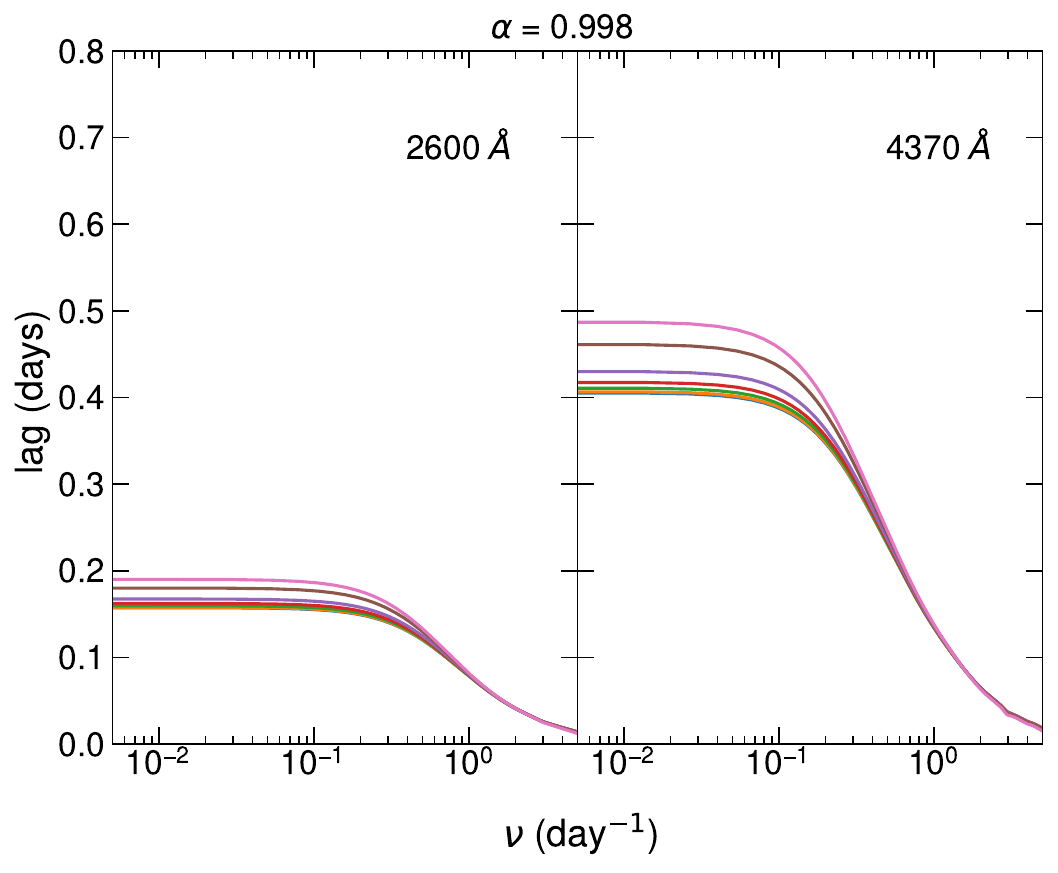}
\caption{Same as in Fig. \ref{fig:phaselags_height} for different values of the X-ray corona power. Here, the power released in the X-ray source is taken from the accretion disk. }    
\label{fig:phaselags_lumin_plus}
\end{figure}

\begin{figure}[h]
\includegraphics[width=0.32\linewidth,height=0.37\linewidth, trim={0 0 0 0}, clip]{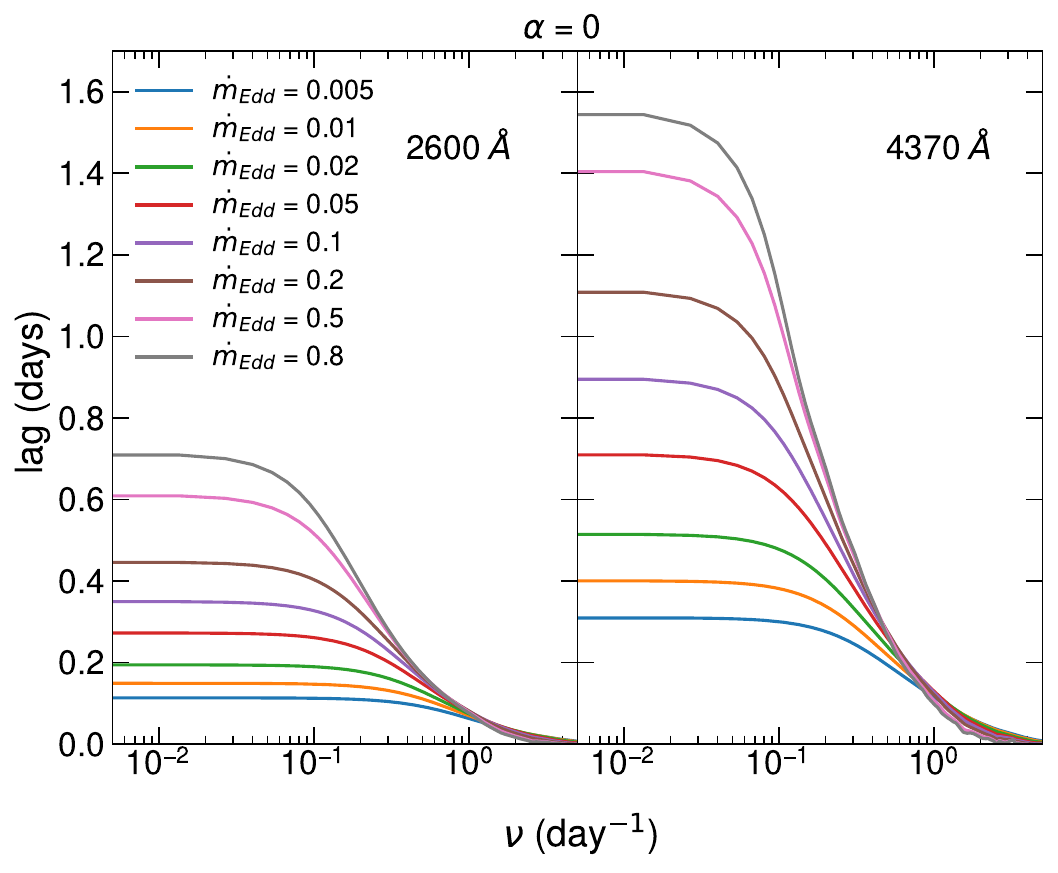}
\includegraphics[width=0.32\linewidth,height=0.37\linewidth, trim={0 0 0 0}, clip]{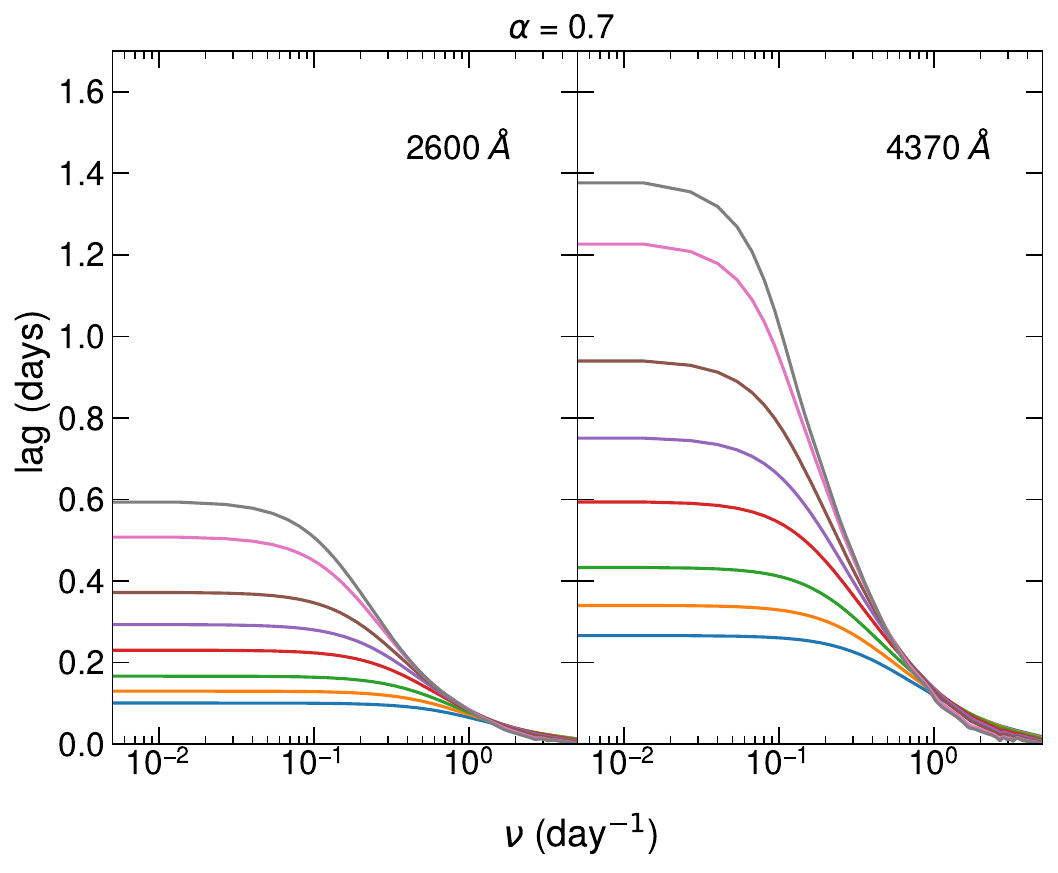}
\includegraphics[width=0.32\linewidth,height=0.37\linewidth, trim={0 0 0 0}, clip]{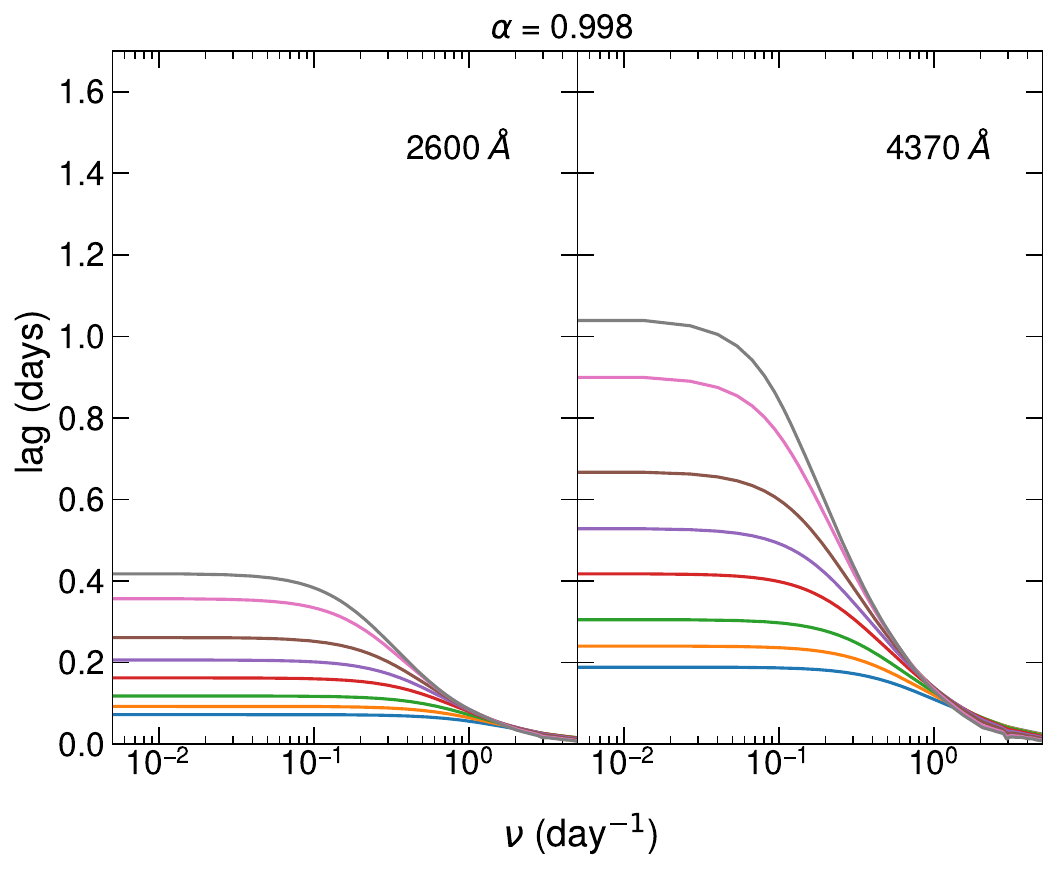}
\caption{Same as in Fig. \ref{fig:phaselags_height} for different values of the accretion rate, measured in units of the Eddington limit.}    
\label{fig:phaselags_mdot}
\end{figure}

\end{document}